\documentclass[11pt]{article} 
\usepackage{graphicx,floatflt,amssymb,epsf,rotate} 
\begin{document} 
\begin{flushright} 
HRI-P-08-12-002 \\
RECAPP-HRI-2008-015\\
CU-Physics/16-2008
\end{flushright} 
 
\vskip 30pt 
 
\begin{center} 
{\Large \bf A note on dimension-5
operators in}
\vskip 5pt 
{\Large \bf GUTs and their impact} \\
\vspace*{1cm} 
\renewcommand{\thefootnote}{\fnsymbol{footnote}} 
{ {\sf Joydeep
Chakrabortty${}^1$\footnote{e-mail: joydeep@hri.res.in}  and}
 {\sf Amitava Raychaudhuri${}^{1,2}$} 
} \\ 
\vspace{10pt} 
   ${}^{1)}$ {\em Harish-Chandra Research Institute,\\
Chhatnag Road, Jhunsi, Allahabad  211 019, India}  

   ${}^{2)}$ {\em Department of Physics, University of Calcutta, \\
92 Acharya Prafulla Chandra Road, Kolkata 700 009, India}  \\

\normalsize 
\end{center} 
 
\begin{abstract} 
\noindent
Quantum gravity or string compactification can lead to effective
dimension-5 operators in Grand Unified Theories which modify the
gauge kinetic terms. We exhaustively discuss the group-theoretic
nature of such operators for the popular SU(5), SO(10), and E(6)
models. In particular, for SU(5) only a Higgs in the 200
representation can help bring the couplings to unification below
the Planck scale and in consistency with proton decay limits
while for a supersymmetric version 24, 75, or 200 representations
are all acceptable. The results also have a direct application in
non-universality of gaugino masses in a class of supersymmetric
models where identical group-theoretic features obtain.

\vskip 5pt 
\noindent 
\texttt{Key Words:~~Grand Unified Theories } 
\end{abstract}

\renewcommand{\thesection}{\Roman{section}} 
\setcounter{footnote}{0} 
\renewcommand{\thefootnote}{\arabic{footnote}} 
 
\section{Introduction}

The remarkable success of electroweak unification has been a
motivation to seek a Grand Unified Theory (GUT) linking together
the strong and electroweak interactions in a framework with
quark-lepton unification \cite{guts}. The merits of this programme need no
underscoring and rightfully it has been attracting continual attention
over several decades. It has all along been also realised that
this is but the penultimate step, unification of all
interactions with gravity being the final objective.

The Standard Model (SM) is based on the gauge group ${\mathcal
G}_{SM} \equiv$ SU(3)$_c$
$\otimes$ SU(2)$_L \otimes$ U(1)$_Y$ which has three independent
couplings $g_3, g_2,$ and  $g_1$. The minimal scheme of grand
unification envisions the placement of quarks and leptons in a
common multiplet of the GUT group and the unification of the
three SM couplings into one unified coupling $g_{GUT}$ at high
energies. The couplings evolve logarithmically with energy and so
the unification, if achieved, is at a high scale of ${\cal
O}(10^{15})$ GeV or more. The current low energy measured values
of the couplings, in fact, are not consistent with unification in
the minimal model. TeV-scale Supersymmetry (SUSY) is one
much-discussed  
remedy for this. Another consequence of grand unification is
the instability of the proton. Proton decay has so far not been
experimentally observed, implying a scale of
grand unification at least an order of magnitude higher than
$10^{15}$ GeV.

A full quantum-theoretic treatment of gravity is not available
currently. Nonetheless, it has been found useful to attempt to
mimic some of its implications on grand unification through
higher dimension effective contributions, suppressed by powers of the
Planck mass, $M_{Pl}$. In a string theory setting, similar effective
operators may also originate from string compactification,
$M_{Pl}$ being then replaced by the compactification scale
$M_c$.

In this work we focus on the corrections to the gauge kinetic
term:
\begin{equation}
\mathcal{L}_{kin}=-\frac{1}{4 c} Tr(F_{\mu\nu} F^{\mu\nu}) .
\label{eq:kin}
\end{equation}
where $F^{\mu\nu} = \Sigma_i \lambda_{i}.F_{i}^{\mu\nu}$ is the gauge field
strength tensor with $\lambda_i$ being the matrix representations
of the generators normalised to
$Tr(\lambda_{i}\lambda_{j})=c~\delta_{ij}$. Conventionally, for
SU($n$) groups the $\lambda_{i}$ are chosen in the fundamental
representation with $c = 1/2$. In the following, we will
often find it convenient to utilise other representations.
 
The lowest order contribution  from quantum gravitational (or
string compactification) effects, which is what we wish to
consider here, is of dimension five and has the form:
\begin{equation}
\mathcal{L}_{dim-5}=-\frac{\eta}{M_{Pl}}\left[\frac{1}{4
c}Tr(F_{\mu\nu}\Phi_{D} F^{\mu\nu})\right]
\label{eq:dim5op}
\end{equation}
where $\Phi_{D}$ denotes the $D$-component Higgs multiplet which
breaks the GUT symmetry and
$\eta$ parametrises the strength of this interaction. In order
for it to be possible to form a gauge invariant of the form in eq.
\ref{eq:dim5op}, $\Phi_D$
can be in any representation included in the symmetric product of
two adjoint representations of the group. For example, in SU(5), 
$(24 \otimes 24)_{sym} = 1 \oplus  24 \oplus 75 \oplus  200 $ and
$\Phi_D$ may be in these representations only.

When  $\Phi_{D}$ develops a 
vacuum expectation value ($vev$) $v_D$, which sets the scale of
grand unification $M_X$, an effective gauge kinetic term is
generated from eq. \ref{eq:dim5op}. Depending on the 
structure of the $vev$, this additional contribution
usually will not be the same for the different subgroups in  ${\mathcal
G}_{SM}$, leading, after an appropriate scaling of the gauge
fields, to an alteration of the gauge coupling unification condition to:
\begin{equation}
g_{1}^{2}(M_{X})(1+\epsilon\delta_1)=g_{2}^{2}(M_{X})(1+\epsilon\delta_2)
=g_{3}^{2}(M_{X})(1+\epsilon\delta_3)  ,
\label{eq:modunif}
\end{equation}
wherein the $\delta_i,
~i=1,2,3$, and $\epsilon = \eta v_{D}/2M_{Pl} \sim {\cal O}
(M_X/M_{Pl})$ arise from eq.  \ref{eq:dim5op}.
Thus, the presence of the dimension-5 terms in the Lagrangian modifies the 
usual boundary conditions on gauge couplings where they are
expected to unify at $M_X$. It is indeed possible that this
change  will be just enough to entail the unification
programme to succeed with the current low energy values of the
coupling constants as a boundary condition. Here, we show that
this is the case for SU(5) GUT.

In this work we work out the consequences of such dimension-5
operators for the unified theories based on SU(5), SO(10), and
E(6). For ordinary SU(5) as well as its SUSY variant, using one-
and two-loop renormalisation group evolution of the gauge
couplings we examine the consistency of the low energy
measurements with grand unification while remaining within
the proton decay restrictions. We also make a brief remark about
the applicability of these results to non-universality of gaugino
masses in a class of SUSY models.

\section{A relook at SU(5) GUTs} \label{sec:su5}

The simplest illustrative example is that of  SU(5) with a
$\Phi_{24}$ scalar. Such a scalar multiplet is
customarily introduced in the theory to spontaneously break SU(5)
$\rightarrow {\mathcal G}_{SM}$. If there is also a dimension-5
term as in eq. \ref{eq:dim5op} involving  $\Phi_{24}$ then gauge
couplings at the high scale get affected \cite{prev}. The $vev$
of this field can be represented as a traceless 5$\times$5
diagonal matrix ($c$ = 1/2):
\begin{equation}
<\Phi_{24}>=\frac{v_{24}}{\sqrt{15}}~diag(1,1,1,-3/2,-3/2) .
\label{vev24}
\end{equation}
The contributions to the $\delta_i$ in eq. \ref{eq:modunif} can
be simply read off from this expression and one finds:
$\delta_1 = \delta_2/3 = -\delta_3/2 = 1/\sqrt{15}$.

For conveniently writing the $vev$ for the 75-dimensional
representation one uses the SU(5) relation: $10 \otimes
\overline{10} = 1 \oplus 24 \oplus 75$. The $vev <\Phi_{75}>$
must  be so chosen that ${\mathcal G}_{SM}$ remains unbroken.
Further, it must be orthogonal to $<\Phi_{24}>$, which too can be
expressed as a $10 \otimes 10$ diagonal matrix.
Under  ${\mathcal G}_{SM}$ the SU(5) 10 =
($\bar{3}$,1)$_{-{4\over3}}$ + (3,2)$_{1\over3}$ + (1,1)$_{2}$.
This allows the identification of the generators of SU(5) in the
10-dimensional representation and, in particular, the U(1)$_Y$
generator  corresponds to $\sqrt{1\over{60}}
diag$(-4,-4,-4,1,1,1,1,1,1,6).  Taking the above into
consideration, $<\Phi_{75}>$ can be expressed as the
traceless 10$\times$10 matrix (c = 3/2):

\begin{equation}
<\Phi_{75}>=\frac{v_{75}}{\sqrt{12}}~diag(1,1,1,-1,-1,-1,-1,-1,-1,3).
\label{vev75}
\end{equation}
This results in $\delta_1 = -5\delta_2/3 = -5\delta_3 = 4/\sqrt{3}$.

Similarly, the relation $15 \otimes \overline{15} = 1 \oplus 24
\oplus 200$ permits the $vev$ for $\Phi_{200}$ to be written as a
(15 $\times$ 15) traceless diagonal matrix (c = 7/2). Noting that under
${\mathcal G}_{SM}$ the 15 of SU(5) is
(6,1)$_{-{4\over3}}$ + (3,2)$_{1\over3}$ + (1,3)$_{2}$ one has:
\begin{equation}
<\Phi_{200}>=\frac{v_{200}}{\sqrt{12}}~diag(1,1,1,1,1,1,-2,-2,-2,-2,
-2,-2,2,2,2),
\label{vev200}
\end{equation}
which yields $\delta_1 = 5\delta_2 = 10\delta_3 =
1/\sqrt{21}$. These results for SU(5) are collected together in
Table  \ref{tab:su5}.

\begin{table}
\begin{center}
\begin{tabular}{|c|c|c|c|} \hline
SU(5)  Representations & $\delta_1$ & $\delta_2$ & $\delta_3$  \\
\hline 
{\bf 24}  & 1/$\sqrt{15}$ & 3/$\sqrt{15}$ & -2/$\sqrt{15}$ \\
\hline
{\bf 75}  &  4/$\sqrt{3}$  & -12/5$\sqrt{3}$ & -4/5$\sqrt{3}$\\
\hline
{\bf 200}  &  1/$\sqrt{21}$   &    1/5$\sqrt{21}$ &   1/10$\sqrt{21}$\\
\hline
\end {tabular}
\caption{Effective contributions to gauge kinetic terms from
different Higgs representations in eq. \ref{eq:dim5op} for SU(5).
(see eq. \ref{eq:modunif}.)}
\label{tab:su5}
\end {center}
\end{table}

\section{$\bf \sin^2\theta_W$, gauge couplings  }

The shift in the gauge couplings as dictated by eq.
\ref{eq:modunif} leaves its mark at low energies through the
Renormalization Group (RG) equations. Moreover, besides the low
energy value of  the weak mixing angle $\sin^2\theta_W$, even its
GUT-level prediction is affected.

The weak mixing angle $\sin^2\theta_W = g'^2/(g^2 + g'^2)$ is
expressed in terms of the SM gauge couplings $g'$ and $g \equiv g_2$. Of
these, the U(1)$_Y$ coupling $g'$  is related to the
coupling $g_1$ arising in a unified theory through $g_1^2 =c^2
g'^2$ where $c^2 = \frac{5}{3}$. In the limit of unification of
all couplings at a GUT-scale, $M_X$, this leads to the prediction
$\sin^2\theta_W(M_X) = 3/8$. Now, due to the modified GUT relationship
of eq. \ref{eq:modunif} one has for the weak mixing angle $\hat{\theta}_W$:
\begin{equation}
\sin^2\hat{\theta}_W(M_X) = {3\over8} + \frac{15}{64} \epsilon (\delta_2
- \delta_1).
\label{eq:xwgut}
\end{equation}
The experimentally determined value of $\sin^2\theta_W$ at low
energies receives further RG-dependent corrections to which we
now turn.

The RG equations governing gauge coupling evolution are:
\begin{equation}
\mu \frac{dg_i}{d\mu} = \beta_i(g_i, g_j),
\label{eq:RG}
\end{equation}
where at two-loop order \cite{beta}
\begin{equation}
\beta_{i}(g_i, g_j)=(16\pi^{2})^{-1}b_{i}g_{i}^{3}
          +(16\pi^{2})^{-2} \sum_{j=1}^{3} b_{ij}g_{j}^{2}g_{i}^{3}.
\end{equation}
$i,j$ = 1,2,3 for U(1)$_Y$, SU(2)$_L$, and SU(3)$_c$, respectively.
The coefficients $b_{i}$ and $b_{ij}$
are:\\
\begin{equation}
 b_{1}={1\over{10}}n_{H}+{4\over3}n_{G};\;\;\;
b_{2}={1\over6}n_{H}+{4\over3}n_{G}-22/3; \;\;\;
b_{3}={4\over3}n_{G}-11,
\label{eq:1loop}
\end{equation}
and
\begin{equation}
b_{ij} =n_{H}\left( \begin{array}{ccc}
9/50 & 9/10 & 0 \\
3/10 & 13/6 & 0 \\
0 & 0 & 0 \\
\end{array}
\right)
\label{sigma}
+n_{G}\left( \begin{array}{ccc}
19/15 & 3/5 & 44/15 \\
1/5 & 49/3 & 4 \\
11/30 & 3/2 & 76/3 \\
\end{array}
\right)
\label{sigma2}
+\left( \begin{array}{ccc}
0 & 0 & 0 \\
0 & -136/3 & 0 \\
0 & 0 & -102 \\
\end{array}
\right).
\label{loop2}
\end{equation}
 and
$n_H$ and $n_G$ are respectively the number of Higgs
doublets and the number of fermion generations in the theory.
The RG equations must satisfy the boundary
conditions set by eq. \ref{eq:modunif} on the $g_{i}^{2}(M_{X})$.

In our numerical analyses below we show the full two-loop RG
equation results. For ease of discussion if only the
lowest order contributions are retained, then in the absence of
dimension-5 operators ($\alpha_i = g_i^2/4\pi$): 
\begin{equation}
 \frac{1}{\alpha_{i}(\mu)} = \frac{1}{\alpha_{i}(M_{X})} +
\frac{2b_{i}}{2\pi} \ln\left[\frac{M_{X}}{\mu}\right], \;\;\; (i=1,2,3).
\end{equation}
$\alpha_i = g_i^2/4\pi$. These equations can be combined to yield:
\begin{equation}
\frac{\alpha}{2\pi}\ln\frac{M_X}{M_Z} = \left[\frac{3}{5 b_1 + 3
b_2 - 8 b_3} \right]
\left\{1 - \frac{8}{3}\frac{\alpha}
{\alpha_3}\right\} ,
\label{eq:Mxstd}
\end{equation}
and  therefrom
\begin{equation}
\sin^2\theta_W (M_Z) = \frac{3}{8} - \frac{15}{8}\left[\frac{b_1 -
b_2}{5 b_1 + 3 b_2 - 8 b_3}\right] 
\left\{ 1 - \frac{8}{3} \frac{\alpha}{\alpha_3} \right\},
\label{eq:xwstd}
\end{equation}
where $\alpha$ -- the fine structure constant -- and $\alpha_3$
are the couplings at the scale $M_Z$.

\begin{table}
\begin{center}
\begin{tabular}{|c|c|c|c|}
\hline
SU(5) & $\epsilon$ & $\epsilon$ & $M_X$ (GeV)  \\
  Representations & (from eq. \ref{eq:sinall})&(using eq. \ref{eq:RG}) & \\
\hline 
{\bf 24}  & 0.087 & 0.088 &5.01$\times 10^{13}$   \\
\hline
{\bf 75}  & -0.048 & -0.045 &4.79  $\times 10^{15}$ \\
\hline
{\bf 200}  & -1.92 & -1.40 & 1.05 $\times 10^{18}$  \\
\hline
\end {tabular}\\
\caption{SU(5) dimension-5 interaction strength $\epsilon$ and the
gauge unification scale,  $M_X$, for different $\Phi$ representations.}
\label{tab:su5uni}
\end {center}
\end{table}

Inclusion of the boundary condition, eq. \ref{eq:modunif},
dictated by the dimension-5 interactions alters eqs. \ref{eq:Mxstd} and
\ref{eq:xwstd} to:
\begin{eqnarray}
\frac{\alpha}{2\pi}\ln\frac{\hat{M}_X}{M_Z} &=& \left[\frac{3}{5 b_1 + 3
b_2 - 8 b_3} \right]
\left\{1 - \frac{8}{3}\frac{\alpha}
{\alpha_3}\right\}
\\ \nonumber
&+& \left(\frac{\epsilon}{5b_1 + 3b_2 - 8b_3}\right)
\left[\frac{-3(8\delta_3 -3\delta_2 - 5\delta_1)b_3}{5b_1 + 3b_2 - 8b_3}
 - (5\delta_1 + 3\delta_2) \frac{\alpha}{\alpha_3}\right]
+{\cal O}(\epsilon^2),
 \label{eq:Mx}
\end{eqnarray}
and 
\begin{eqnarray}
\label{eq:xw}
\sin^2\hat{\theta}_W (M_Z) &=& \frac{3(1 + \epsilon\delta_2)}{8 +
\epsilon(3\delta_2 + 5\delta_1)} \\ \nonumber
&-& \left(\frac{5(1+
\epsilon\delta_1)(1+ \epsilon\delta_2)}{8+\epsilon(3\delta_2 +
5\delta_1)}\right) \left[\frac{b_1}{1 + \epsilon\delta_1} -
\frac{b_2}{1 + \epsilon\delta_2}\right]
\frac{3(1 + \epsilon\delta_3)  - [8 +
\epsilon(3\delta_2 + 5\delta_1)][\alpha/\alpha_3]}{(5b_1 + 3 b_2)(1 +
\epsilon\delta_3) - [8 + \epsilon(3\delta_2 + 5\delta_1)]b_3} ,
\end{eqnarray}
which  reduces to eq. \ref{eq:xwstd} in the
appropriate limit. In fact,   
\begin{eqnarray}
 \label{eq:sinall}
\sin^2\hat{\theta}_W (M_Z) &=& \sin^2\theta_W (M_Z) 
\\
\nonumber
&-& \epsilon \left[ \frac{5[\delta_1(b_3 - b_2) + \delta_2(b_1 - b_3)  
+ \delta_3(b_2 - b_1)]}{(5b_1 + 3b_2 - 8b_3)^2}\left\{3b_3 -
(5b_1 + 3b_2)\frac{\alpha}{\alpha_3}\right\}
\right]
+{\cal O}(\epsilon^2).
\end{eqnarray}
The first term on the r.h.s. of eq. \ref{eq:sinall} is fixed
from eq. \ref{eq:xwstd}. From the measured value of
$\sin^2\theta_W$ \cite{pdg} one can extract the value of
$\epsilon$. These are presented for the different $\Phi$
representations in Table \ref{tab:su5uni}.

\begin{center}
\begin{figure}[thb]
\hskip 1.0cm
\includegraphics[width=4cm,height=4.0cm,angle=0]{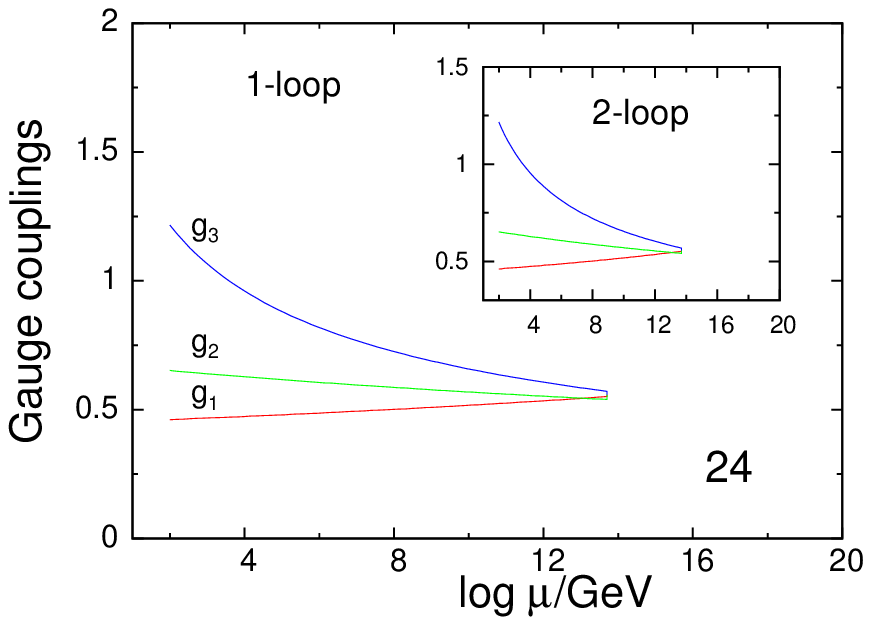}
\hskip 0.5cm
\includegraphics[width=4cm,height=4.0cm,angle=0]{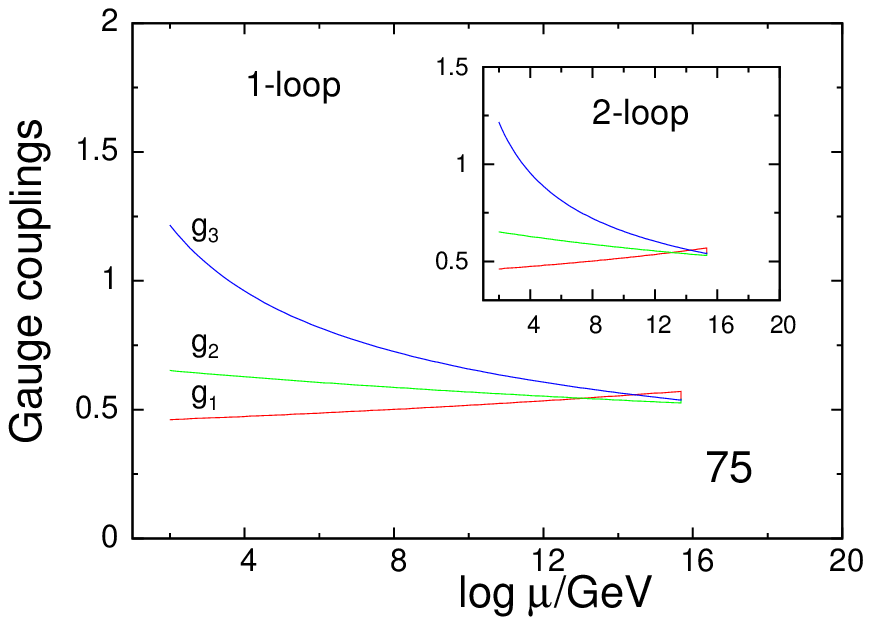}
\hskip 0.5cm
\includegraphics[width=4cm,height=4.0cm,angle=0]{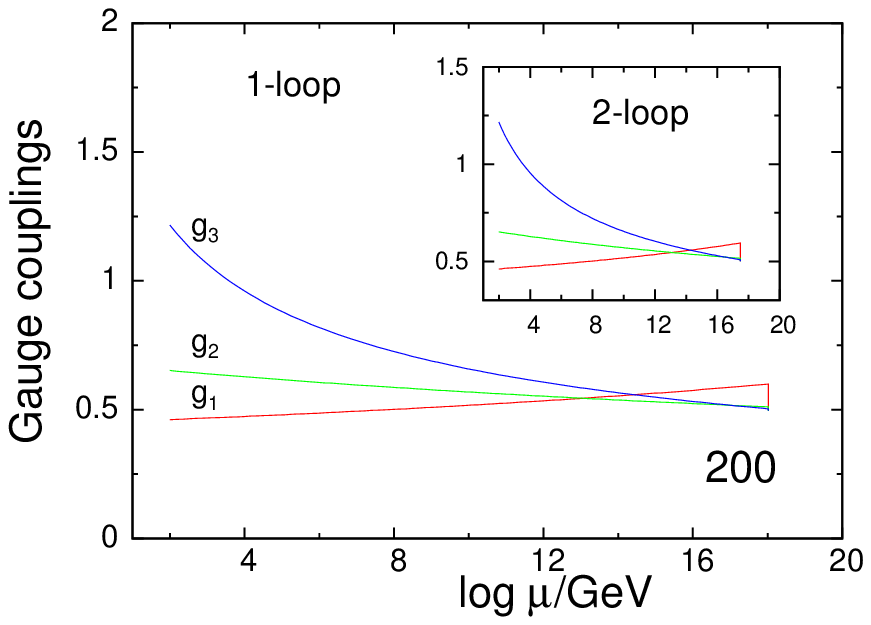}
\caption{\sf \small {The evolution of gauge coupling constants for
different choices of $\Phi$ for SU(5) GUTs: 
$\Phi_{24}$ (left),  $\Phi_{75}$ (centre),  $\Phi_{200}$ (right).
In the inset the results for two-loop evolution are shown.}}
\label{f:evo}
\end{figure}
\end{center}

These ${\cal O}(\epsilon)$ one-loop analytic results can be
cross-checked using the full RG equations in eq. \ref{eq:RG} with
$n_G = 3$ and $n_H = 1$.  Using the low energy ($\sim M_Z$)
measured values \cite{pdg}, $\sin^2\theta_W$ = 0.231 19(14) and
$\alpha_3$ = 0.11 76(20), the RG equations can be numerically
integrated. The scale $M_X$ is fixed through the requirement that
the modified unification condition, eq. \ref{eq:modunif}, is
satisfied there. From this analysis one can determine $\epsilon$
and $M_X$. The conclusions from one-loop RG running are shown in
Table \ref{tab:su5uni} and Fig. \ref{f:evo}.

The two-loop results,  shown as insets in Fig.
\ref{f:evo},  incorporate the proper matching conditions
\cite{hall2lp} as well as eq. \ref{eq:modunif} at $M_X$, namely,  
\begin{equation}
\frac{1}{\alpha_i(M_X)(1 + \epsilon \delta_i)} - \frac{C_i}{12
\pi} = {\rm ~constant, ~independent ~of} ~i \,\,\,\,\,{\rm for} ~i=1,2,3, 
\end{equation}
where $C_i$ is the quadratic Casimir for the $i$-th subgroup. It
is noteworthy that the results are not significantly affected and
the coupling constants still unify. The unification scales, $M_X$,
are found to be 5.01 $\times 10^{13}$, 2.09
$\times 10^{15}$, and 3.02 $\times 10^{17}$ GeV respectively for
$\Phi_{24}$, $\Phi_{75}$, and $\Phi_{200}$.  Though  unification
is achieved within the Planck scale for all three choices, for
$\Phi_{24}$ and  $\Phi_{75}$ the results are not consistent with
the existing limits from proton decay. Thus only a 5-dimensional
operator with $\Phi_{200}$ yields a viable solution.

In \cite{utpal}, noting that the dimension-5 operator in eq.
\ref{eq:dim5op} with $\Phi_{24}$ cannot by itself provide
satisfactory gauge unification, it has been proposed that
including gravitational contributions in the beta functions can
help ameliorate this problem. Alternatively, within SUSY SU(5) it
has been argued  in \cite{calmet} that one-loop (as well as
two-loop) RG evolution with $\Phi_{24}$-driven boundary
conditions in eq. \ref{eq:modunif} can yield satisfactory
unification solutions provided the possible modification of the
Planck scale itself due to the large number of GUT fields is
given consideration.

\section{SO(10) and E(6)}

SU(5) is admittedly the GUT model which has been examined most in
the literature by far. Nonetheless, there are also strong
motivations for going to larger GUT groups like SO(10) and E(6).
The minimal SU(5) model, as noted earlier, is in conflict with
proton decay measurements.  Further, the larger gauge groups
offer the possibility of left-right symmetry. Here, we briefly
discuss the effect of dimension-5 interactions, as in eq.
\ref{eq:dim5op}, on SO(10) and E(6) models. The detailed analysis
of RG evolution of the gauge couplings for these cases is
not presented in this paper
\cite{cr2}.  

\subsection{SO(10) GUT}
SO(10) \cite{so10} is now the widely preferred model for grand unification,
offering the option of descending to ${\mathcal G}_{SM}$ through
a left-right symmetric route \cite{lrs} -- the intermediate
Pati-Salam ${\mathcal G}_{PS} \equiv$ SU(4)$_c \otimes$ SU(2)$_L
\otimes$ SU(2)$_R $ group -- or {\em via} an  SU(5) $\otimes$ U(1)
theory. The latter option will not give any novel feature
beyond what has already been discussed in the context of SU(5)
GUTs. So, here we consider the breaking chain SO(10)
$\rightarrow {\mathcal G}_{PS} \rightarrow {\mathcal G}_{SM}$.

In SO(10), one usually requires $g_{4c} = g_{2L} = g_{2R}$ at the
unification scale. The presence of any dimension-5 effective
interactions, of the form of eq. \ref{eq:dim5op}, will affect
this relation introducing corrections such that
\begin{equation}
g_{4c}^{2}(M_{X})(1+\epsilon\delta_{4c}) =
g_{2L}^{2}(M_{X})(1+\epsilon\delta_{2L}) =
g_{2R}^{2}(M_{X})(1+\epsilon\delta_{2R}).
\label{eq:modunif2}
\end{equation}

The adjoint representation 45 of SO(10) satisfies $(45 \otimes
45)_{sym} = 1 \oplus 54 \oplus 210 \oplus  770 $ implying that
$\Phi_D$ can be chosen only in the 54, 210, and 770-dimensional
representations. The $vev$ $<\Phi_D >$ must ensure that ${\mathcal
G}_{PS}$ is unbroken.

Using the SO(10) relation $(10 \otimes 10) = 1 \oplus  45 \oplus
54$ one can see that $<\Phi_{54}>$ can be expressed as a 10
$\times$ 10 diagonal traceless matrix. Under  SU(4)$_c \otimes$
SU(2)$_L \otimes$ SU(2)$_R $, 10 $\equiv$ (1,2,2) + (6,1,1).  Thus
(c=1)
\begin{equation}
<\Phi_{54}>=\frac{v_{54}}{2\sqrt{15}}~diag(3,3,3,3,-2,-2,-2,-2,-2,-2).
\label{ve54}
\end{equation}
This leads to $\delta_{4c} = -{1\over{\sqrt{15}}}$ and
$\delta_{2L} = \delta_{2R} = {3\over{2\sqrt{15}}}$.  Notice that
this correction to unification, through eq.
\ref{eq:modunif2}, ensures that $g_{2L}(M_X) = g_{2R}(M_X)$, i.e.,
D-parity \cite{dpar} is preserved.
\begin{table}
\begin{center}
\begin{tabular}{|c|c|c|c|} \hline
SO(10)  Representations & $\delta_{4c}$ & $\delta_{2L}$ & $\delta_{2R}$  \\
\hline 
{\bf 54}  & -1/$\sqrt{15}$ & 3/$2\sqrt{15}$ & 3/$2\sqrt{15}$\\
\hline
{\bf 210}  &  0 & 1/$\sqrt2$ & -1/$\sqrt2$\\
\hline
{\bf 770}  &   2/$3\sqrt5$ &  5/$3\sqrt5$   & 5/$3\sqrt5$\\
\hline
\end {tabular}
\caption{Effective contributions to gauge kinetic terms from
different Higgs representations in eq. \ref{eq:dim5op} for SO(10).
(see eq. \ref{eq:modunif}.)}
\label{tab:so10}
\end {center}
\end{table}

In SO(10) $(\overline{16} \otimes 16) = 1 \oplus  45 \oplus 210 $;
so one
can represent $<\Phi_{210}>$ as a 16-dimensional traceless
diagonal matrix. Since 
 16 $\equiv$ (4,2,1) + ($\bar4$,1,2) one can readily identify (c=2)
\begin{equation}
<\Phi_{210}>=\frac{v_{210}}{2\sqrt{2}}~diag(1,1,1,1,1,1,1,1,
-1,-1,-1,-1,-1,-1,-1,-1),
\label{vev210}
\end{equation}
yielding $\delta_{4c} = $0 and $\delta_{2L} = -\delta_{2R} =
{1\over\sqrt2}$. It is noteworthy that D-parity is broken through
$<\Phi_{210}>$ and $g_{2L}(M_X) \neq g_{2R}(M_X)$ though SU(2)$_L
\otimes$ SU(2)$_R$ remains unbroken at $M_X$. The effect of dimension-5
interactions arising from $\Phi_{54}$ and $\Phi_{210}$ 
on gauge coupling unification has been examined in the
literature \cite{appl}.

Finally, we turn to the last possibility for SO(10), namely
$\Phi_{770}$. Since $(45 \otimes 45)_{sym} = 1 \oplus  54 \oplus
210 \oplus  770 $ with  45 $\equiv$ (15,1,1) + (1,3,1) + (1,1,3)+
(6,2,2) one can write the $vev$ in terms of a 45 $\times$ 45
diagonal traceless matrix. The $<\Phi_{54}>$ and $<\Phi_{210}>$
can also be written in a similar form and care must be taken to
ensure that $<\Phi_{770}>$ is orthogonal to them. In this manner
one arrives at (c=8):
\begin{equation}
<\Phi_{770}>=\frac{v_{770}}{\sqrt{180}}~diag(\underbrace{-4,\ldots,-4}_{15
~entries},\underbrace{-10,\ldots,-10}_{3+3
~entries},\underbrace{5,\ldots,5}_{24
~entries}).
\label{vev770}
\end{equation}

From this one finds  $\delta_{4c} = {2\over{3\sqrt{5}}}$ and
$\delta_{2L} = \delta_{2R} = {5\over{3\sqrt{5}}}$.

All the results for SO(10) are collected together in Table \ref{tab:so10}.

\subsection{E(6) GUT}
The exceptional group E(6) has been proposed as a viable GUT
alternative \cite{e6}. It offers SU(3)$_c \otimes$ SU(3)$_L
\otimes$ SU(3)$_R$ as a subgroup besides SO(10) $\otimes$ U(1). 

For E(6) the adjoint representation is 78-dimensional.
Noting that $(78 \otimes 78)_{sym} = 1 \oplus  650 \oplus 2430$
it is clear that $\Phi_D$ can be only in the 650 and
2430-dimensional representations in this case.

In E(6), $(\overline{27} \otimes 27) = 1 \oplus  78 \oplus 650 $
and under SU(3)$_c \otimes$ SU(3)$_L \otimes$ SU(3)$_R$
27 =  (1,$\bar3$,3) + (3,1,$\bar3$) + ($\bar3$,3,1). Therefore one
can write $<\Phi_{650}>$ as a 27 $\times$ 27 diagonal traceless
matrix (c=3). Further, the 650 representation has two fields
which are singlet under SU(3)$_c \otimes$ SU(3)$_L \otimes$
SU(3)$_R$. Needless to say, any one of these fields or linear
combinations thereof may be chosen to break the symmetry. In
particular, two linear combinations may be identified which
respect $\delta_{3L} = \pm \delta_{3R}$. These are:
\begin{equation}
<\Phi_{650}>=\frac{v_{650}}{\sqrt{18}}~diag(\underbrace{-2,\ldots,-2}_{9
~entries},\underbrace{1,\ldots,1}_{9
~entries},\underbrace{1,\ldots,1}_{9~entries}).
\label{vev650}
\end{equation}
This results in $\delta_{3c} = -{1\over{\sqrt{2}}}$ and
$\delta_{3L} = \delta_{3R} = {1\over{2\sqrt{2}}}$,
and
\begin{equation}
<\Phi'_{650}>=\frac{v'_{650}}{\sqrt{6}}~diag(\underbrace{0,\ldots,0}_{9
~entries},\underbrace{1,\ldots,1}_{9
~entries},\underbrace{-1,\ldots,-1}_{9~entries}).
\label{vev650'}
\end{equation}
From this $\delta_{3c} = $0 and
$\delta_{3L} = -\delta_{3R} = {3\over{2\sqrt{6}}}$.

Finally, using 
78 = (8,1,1) + (1,8,1) + (1,1,8) + (3,3,$\bar3$) +
($\bar3$,$\bar3$,3) and maintaining consistency with
$<\Phi_{650}>$ and $<\Phi'_{650}>$
one can write (c=12)
\begin{equation}
<\Phi_{2430}>=\frac{v_{2430}}{3\sqrt{26}}~diag(\underbrace{9,\ldots,9}_{8
~entries},\underbrace{9,\ldots,9}_{8
~entries},\underbrace{9,\ldots,9}_{8
~entries},\underbrace{-4,\ldots,-4}_{27
~entries},\underbrace{-4,\ldots,-4}_{27
~entries}).
\label{vev2430}
\end{equation}
One can readily read off  $\delta_{3c} = 
\delta_{3L} = \delta_{3R} = -{3\over{\sqrt{26}}}$.

In Table \ref{tab:e6} we collect the findings for the different
representations of E(6).

\begin{table}
\begin{center}
\begin{tabular}{|c|c|c|c|} \hline
E(6)  Representations & $\delta_{3c}$ & $\delta_{3L}$ & $\delta_{3R}$  \\
\hline 
{\bf 650}  & - 1/$\sqrt2$ &  1/2$\sqrt2$ &  1/2$\sqrt2$ \\
\hline
{\bf 650$^\prime$}  & 0 &  3/2$\sqrt6$ &  -3/2$\sqrt6$ \\
\hline
{\bf 2430}  &  -${3\over{\sqrt{26}}}$ & -${3\over{\sqrt{26}}}$ &
-${3\over{\sqrt{26}}}$\\
\hline
\end {tabular}
\caption{Effective contributions  to gauge kinetic terms from
different Higgs representations in eq. \ref{eq:dim5op} for E(6).
(see eq. \ref{eq:modunif}.) Note that there are two SU(3)$_c
\otimes$ SU(3)$_L \otimes$ SU(3)$_R$ singlet directions in 650.}
\label{tab:e6}
\end {center}
\end{table}

\section{Supersymmetric GUTs}

\subsection{Unification, neutrino mass}
In a supersymmetric theory, the fermions, Higgs scalars, and
gauge bosons of the GUT model are endowed with superpartners
constituting chiral and vector supermultiplets. It is well known
that gauge coupling unification is possible if SUSY is manifested
at the TeV scale \cite{susygut}. If dimension-5 interactions are
also present then that will further affect this unification. In
fact, it was shown within SUSY SU(5) that if  the $\delta_i \;
(i=1,2,3)$ in eq.
\ref{eq:modunif} are fixed as determined (see Table
\ref{tab:su5}) by the 24-dimensional
representation \cite{susy5e} or permitted to vary arbitrarily
\cite{susy5d} then unification, at the one-loop level, is always
possible. 

Here we perform a one-loop as well as a two-loop analysis. Above
the SUSY scale (chosen as 1 TeV) this entails the replacement of
eqs. \ref{eq:1loop} and \ref{loop2}  by (for $n_G = 3, \; n_H = 2$)
\begin{equation}
 b_{1}={33\over 5};\;\;\;
b_{2}={1}; \;\;\;
b_{3}={-3},
\label{eq:1loops}
\end{equation}
and
\begin{equation}
b_{ij} =\left( \begin{array}{ccc}
199/25 & 27/5 & 88/5 \\
9/5 & 25 & 24 \\
11/5 & 9 & 14 \\
\end{array}
\right).
\label{loop2s}
\end{equation}
We find that unification is allowed for all three
choices of $\Phi$ -- namely, 24, 75, and 200 -- when the
$\delta_i \; (i=1,2,3)$ are appropriately identified. The results
are presented in Table \ref{tab:su5susy}.  Unlike 
the non-SUSY alternative in Table
\ref{tab:su5uni}, now for every case one gets $M_X \sim 10^{16}$
GeV.  In line with expectation, the size of $\epsilon$ is reduced
in this SUSY case as the couplings tend to unify even without
these interactions.  The trend of agreement between the one-loop
and two-loop results is gratifying.

\begin{table}
\begin{center}
\begin{tabular}{|c|c|c|c|c|}
\hline
SU(5) & \multicolumn{2}{|c|}{1 loop} & 
\multicolumn{2}{|c|}{2 loop}\\ \cline{2-5}
Representations & $\epsilon$ & $M_X$ (GeV)& $\epsilon$ & $M_X$ (GeV) \\
\hline 
{\bf 24}  & 0.017 &1.10$\times 10^{16}$& -0.003 &1.38$\times 10^{16}$   \\
\hline
{\bf 75}  & -0.007 &1.92  $\times 10^{16}$& 0.001 &1.24$\times 10^{16}$ \\
\hline
{\bf 200}  & -0.204 & 3.16 $\times 10^{16}$& 0.071 &1.10$\times 10^{16}$  \\
\hline
\end {tabular}\\
\caption{SU(5) dimension-5 interaction strength $\epsilon$ and the
gauge unification scale,  $M_X$, for different $\Phi$
representations in a supersymmetric theory.}
\label{tab:su5susy}
\end {center}
\end{table}

The degenerate fermionic SUSY partner of $\Phi$ might play a role
in the generation of realistic neutrino masses through the
see-saw mechanism. Continuing with the SU(5) model for the
purpose of illustration, if $\Phi$ is in the 24 representation
then its fermion partner multiplet contains fields transforming
as (1,1,0) and (1,3,0) under SU(3)$_c
\otimes$SU(2)$_L \otimes$U(1)$_Y$. Through a Yukawa coupling of
the form 5$_{\rm F}$24$_{\rm F}\overline{5}_{\rm H}$ they can
serve as the heavy exchanged fermionic mode of type I
\cite{seesaw1} and type III \cite{seesaw3} see-saw models,
respectively\footnote{It is true that the Yukawa coupling of
these superpartner fermions with an ordinary fermion and a Higgs
scalar -- as required in the see-saw structure --  will be
R-parity non-conserving. This does not contradict experimental
results since the fermions in the 24 multiplet are all superheavy.}.
These fermions reside at the $M_X$ scale and will lead to
neutrino masses typically at the $10^{-2} - 10^{-3}$ eV level. In
fact, the above Yukawa coupling will result in
simultaneous type I and type III see-saw contributions which
together will lead to an enhancement.  A similar situation
obtains if $\Phi$ is chosen in the 200 representation. In the 75
of SU(5) the SU(3)$_c\otimes$SU(2)$_L \otimes$U(1)$_Y$ (1,3,0)
field is not present. So only type I see-saw is feasible for this
alternative. For SO(10) and E(6) SUSY GUTs,  over and above these
options, there is even room for
the type II see-saw \cite{seesaw2} through heavy scalars
transforming as (1,3,0).

\subsection{Non-universality of gaugino masses}
The results presented in the earlier sections have a direct application --
specifically in the generation of non-universal gaugino masses --
in SUSY models which  emerge from a GUT where supersymmetry
breaking and GUT symmetry breaking are tied together. These arise
from the gauge kinetic term which can be schematically written as
\begin{equation}
{\cal L} = \int d^2\theta f_{\alpha \beta}(\Phi) ~W^\alpha
W^\beta + h.c. ,
\label{susy1}
\end{equation}
where $f_{\alpha \beta}$ is a function of the chiral superfield
$\Phi$ whose scalar component is responsible for the GUT symmetry
breaking. $f_{\alpha \beta}$ is symmetric in the GUT gauge
indices $\alpha, \beta$ and $W^\alpha$ are the GUT gauge
superfields.  When the F-component of the chiral superfield,
$F_\Phi$, gets a non-zero $vev$ at the GUT scale, the gauginos
$\lambda^\alpha$ develop an effective mass term
\begin{equation}
{\cal L_{\rm mass}} \propto \frac{<F_\Phi>_{\alpha\beta}}{M} ~\lambda^\alpha
\lambda^\beta ,
\label{susy2}
\end{equation}
where $M$ is the mass scale of the GUT symmetry breaking. This
scenario, including detailed phenomenological implications, has
been widely discussed in the context of SUSY SU(5)
\cite{susy5e, susy5d}. Some cases have also been examined for SUSY SO(10)
\cite{susy10}.

It is obvious that the group-theoretic structures of eqs. \ref{eq:dim5op}
and \ref{susy2} are identical. Hence the results discussed in the
previous sections can be taken over {\em mutatis mutandis}. For
example, for SU(5) breaking to SU(3)$_c \otimes $SU(2)$_L \otimes $U(1)$_Y$ we
should get for the gaugino masses $M_i$:
\begin{equation}
M_{3c}:M_{2L}:M_Y = \delta_3:\delta_2:\delta_1.
\end{equation} 
This is to be compared with eq. \ref{eq:modunif}. The $\delta_{1,2,3}$
can be found in Table \ref{tab:su5}. The non-universality of
gaugino masses has far-reaching implications for SUSY
phenomenology \cite{susy5_2}.

Similarly, for  SO(10) $\rightarrow$  SU(4)$_c \otimes $SU(2)$_L
\otimes$ SU(2)$_R$ one has
\begin{equation}
M_{4c}:M_{2L}:M_{2R} = \delta_{4c}:\delta_{2L}:\delta_{2R},
\end{equation} 
where $\delta_{4c,2L,2R}$ can be found in Table \ref{tab:so10}.
Of course, at an intermediate scale there is the further breaking
of SU(4)$_c \otimes $SU(2)$_R$ to SU(3)$_c \otimes $ U(1)$_Y$ and
for phenomenology at low energies the SU(3)$_c$, SU(2)$_L$, and
U(1)$_Y$ gauginos are of relevance \cite{susy10_2}.

Finally, for E(6) breaking to SU(3)$_c \otimes$ SU(3)$_L \otimes$
SU(3)$_R$:
\begin{equation}
M_{3c}:M_{3L}:M_{3R} = \delta_{3c}:\delta_{3L}:\delta_{3R}.
\end{equation} 
The deviation from universality for different Higgs
representations is controlled by the  $\delta_{3c,3L,3R}$ which
are listed in Table \ref{tab:e6}.

\section{Conclusions}

Non-perturbative effects arising from quantum gravity or string
compactification can be mimicked through higher dimensional
non-renormalisable interactions. We have considered one class of
such dimension-5 interactions which modify the unification
condition of gauge coupling constants in grand unified theories.
For SU(5), SO(10), and E(6) GUTs we have exhaustively  worked out
their implications for gauge coupling unification conditions. For
the case of SU(5), we have shown how low energy physics can
constrain these interaction strengths and the manner in which
this would affect the unification scale. We also briefly 
examine the status of gauge unification in an SU(5) SUSY GUT
model with the dimension-5 interactions and discuss a possible
route to generate realistic  neutrino masses through the see-saw
mechanism.  A corollary of the exercise is an application to SUSY
models where non-universal gaugino mass relations obtain from a
similar group-theoretic structure.

\vskip 20pt

{\bf Acknowledgements} This research has been supported by funds
from the XIth Plan   RECAPP and `Neutrino Physics' projects at
HRI. JC is thankful to Nishita Desai for discussions about the
numerical work. Use of the HRI cluster computational facility is
gratefully acknowledged.

\end{document}